\documentclass[twocolumn,showpacs,preprintnumbers,amsmath,amssymb,nofootinbib,floatfix]{revtex4-1}

\usepackage{graphicx}

\def\Journal#1#2#3#4{{#1} {\bf #2}, #3 (#4)}

\def\APJ{Astrophys. J.}

\def\EPJC{Eur. Phys. J. C}

\def\JHEP{JHEP}

\def\JETPUSSR{JETP (USSR)}
\def\JPG{J. Phys. G}

\def\NPB{Nucl. Phys. B}

\def\NPBSUPPL{Nucl. Phys. B. Proc. Suppl.}

\def\PLB{Phys. Lett. B}

\def\PLBOLD{Phys. Lett.}
\def\PAN{Phys. Atom. Nucl.}
\def\PRL{Phys. Rev. Lett.}
\def\PRD{Phys. Rev. D}

\def\PTP{Prog. Theor. Phys.}
\def\RMP{Rev. Mod. Phys.}

\def\SCIENCE{Science}

\def\ZETP{Zh. Eksp. Teor. Piz.}

\begin{document}

\title{Formulae for flavour neutrino masses and its application to texture two zeros}

\author{Teruyuki Kitabayashi}
\email{teruyuki@keyaki.cc.u-tokai.ac.jp}
\author{Masaki Yasu\`{e}}
\email{yasue@keyaki.cc.u-tokai.ac.jp}
\affiliation{Department of Physics, Tokai University, 4-1-1 Kitakaname, Hiratsuka, Kanagawa, 259-1292, Japan}
\date{\today}

\begin{abstract}
We demonstrate the usefulness of flavour neutrino masses expressed in terms of $M_{ee},M_{e\mu}$ and $M_{e\tau}$. The analytical expressions for the flavour neutrino masses, mass eigenstates and physical CP-violating Majorana phases for texture two zeros are obtained exactly. 
\end{abstract}

\maketitle

\section{Introduction}
\label{sec:introduction}
The atmospheric neutrino oscillations have been experimentally confirmed by the Super-Kamiokande collaboration \cite{atmospheric}. A similar oscillation phenomenon, the solar neutrino oscillations, has been long suggested \cite{solarold} and have been finally confirmed by various collaborations \cite{solar}.  

Theoretically, the neutrino oscillations are realised if neutrinos have different masses and can be explained by mixings of three flavour neutrinos $(\nu_e,\nu_\mu,\nu_\tau)$. These mixings are well described by a unitary matrix $U$ \cite{PMNS} involving three mixing angles $\theta_{12},\theta_{23},\theta_{13}$, which converts three neutrino mass eigenstates $(\nu_1,\nu_2,\nu_3)$ with masses $(m_1,m_2,m_3)$ into $(\nu_e,\nu_\mu,\nu_\tau)$. Furthermore, leptonic CP violation is induced if $U$ contains phases which are given by one CP-violating Dirac phase $\delta$ and two CP-violating Majorana phases $\alpha_2,\alpha_3$ \cite{CPViolationOrg}. 

There have been various discussions of flavour neutrino mass matrix to ensure the appearance of the observed neutrino mixings and masses \cite{ReviewOfMixingMatrix}. The flavour neutrino mass matrix $M$ is parameterized by
\begin{eqnarray}
M &=& 
\left( 
\begin{array}{*{20}{c}}
M_{ee} & M_{e\mu} & M_{e\tau} \\
M_{e\mu} & M_{\mu\mu} & M_{\mu\tau} \\
M_{e\tau} & M_{\mu\tau} & M_{\tau\tau} \\
\end{array}
\right),
 \label{Eq:M_nu}
\end{eqnarray}
where ${\rm diag}(m_1, m_2, m_3)= U^T M U$. The elements of the mass matrix, $M_{ij}$ ($i,j=e,\mu,\tau$), are functions of the mixing angles, mass eigenstates and CP phases :
\begin{eqnarray}
M_{ij} = f(\theta_{12},\theta_{23}, \theta_{13}, m_1,m_2,m_3, \delta, \alpha_2, \alpha_3).
\end{eqnarray}

Our recent discussions \cite{KitabayashiYasue2015arXiv} have found that $M_{\mu\mu},M_{\tau\tau},M_{\mu\tau}$ as well as $m_1,m_2,m_3$ are expressed in terms of $M_{ee},M_{e\mu},M_{e\tau}$ as follows:
\begin{eqnarray}
M_{\mu\mu,\tau\tau,\mu\tau} = f(M_{ee},M_{e\mu},M_{e\tau},\theta_{12},\theta_{23}, \theta_{13},\delta),
\label{Eq:M_func}
\end{eqnarray}
and
\begin{eqnarray}
m_{1,2,3} = f(M_{ee},M_{e\mu},M_{e\tau}, \theta_{12},\theta_{23}, \theta_{13}, \delta, \alpha_2, \alpha_3).
\label{Eq:m_func}
\end{eqnarray}
Various specific textures of flavour neutrino mass matrices for the maximal CP violation and the maximal atmospheric neutrino mixing have been obtained by using the relation of Eq.(\ref{Eq:M_func}) \cite{KitabayashiYasue2015arXiv}.

In this article, we would like to remind you about the useful formulae for the flavour neutrino masses expressed in terms of $M_{ee},M_{e\mu}, M_{e\tau}$. In Sec.\ref{sec:formulae}, we show the useful formulae. In Sec.\ref{sec:texture_two_zeros}, to demonstrate the usefulness of flavour neutrino masses expressed in terms of $M_{ee},M_{e\mu}$ and $M_{e\tau}$, we use the formulae to find exact analytical expressions for flavour neutrino masses, mass eigenstates and CP-violating Majorana phases in the texture two zeros. To confirm results of our discussions more concretely, we study the following three subjects based on numerical and semi-analytical calculations: (1) the consistency between our results and conclusions obtained from other papers, (2) the upper limit of the effective Majorana neutrino mass for the neutrino less double beta decay for maximal Dirac CP-violation and (3) the dependence of the CP-violating Majorana phases on the CP-violating Dirac phase. The final section Sec.\ref{sec:summary} is devoted to summary.

\section{\label{sec:formulae} Formulae}
\subsection{Formulae}
We summarise the useful formulae found in Ref \cite{KitabayashiYasue2015arXiv}. We adopt the standard parameterization of the unitary matrix $U=U_0 K$ \cite{PDG} with $U_0=U_0^{\rm PDG}$ where
\begin{eqnarray}
U_0^{\rm PDG} &=&
\left( 
\begin{array}{*{20}{c}}
1 & 0 & 0 \\
0 & c_{23} & s_{23} \\
0 & -s_{23} & c_{23} \\
\end{array}
\right)
\left( 
\begin{array}{*{20}{c}}
c_{13} & 0 & s_{13}e^{-i\delta} \\
0 & 1 & 0 \\
-s_{13}e^{i\delta} & 0 & c_{13} \\
\end{array}
\right)
\nonumber \\
&& \times
\left( 
\begin{array}{*{20}{c}}
c_{12} & s_{12} & 0 \\
-s_{12} & c_{12} & 0 \\
0 & 0 & 1 \\
\end{array}
\right),
\label{Eq:U_0}
\end{eqnarray}
and
\begin{eqnarray}
K &=& {\rm diag}(e^{i\phi_1/2}, e^{i\phi_2/2}, e^{i\phi_3/2}),
\label{Eq:K}
\end{eqnarray}
where $c_{ij}=\cos\theta_{ij}$ and $s_{ij}=\sin\theta_{ij}$ (as well as $t_{ij}=\tan\theta_{ij}$) and $\theta_{ij}$ represents the $\nu_i$-$\nu_j$ mixing angle ($i,j$=1,2,3). The phases $\phi_{1,2,3}$ stand for the Majorana phases. The physical Majorana CP-violation is specified by two combinations made of $\phi_1,\phi_2,\phi_3$ such as $\alpha_i = \phi_i - \phi_1$ in place of $\phi_i$ in $K$ \cite{CPViolationOrg}: 
\begin{eqnarray}
K = {\rm diag}(1, e^{i\alpha_2/2}, e^{i\alpha_3/2}).
\label{Eq:K_PDG}
\end{eqnarray}

Defining that

\begin{eqnarray}
{M_ + } &=& {s_{23}}{M_{e\mu }} + {c_{23}}{M_{e\tau }}, \nonumber \\
{M_ - } &=& {c_{23}}{M_{e\mu }} - {s_{23}}{M_{e\tau }},
\end{eqnarray}
and 
\begin{eqnarray}M_{\mu\tau }^{\left( 0 \right)} &=& -\left( \frac{1}{{{c_{13}}\tan 2{\theta_{12}}}} + \frac{t_{13}e^{ - i\delta}}{\tan 2\theta_{23}} \right)M_-  \nonumber \\
&& + \left( \frac{{{e^{ - i{\delta}}}}}{{\tan 2{\theta_{13}}}} + \frac{1}{2}t_{13}e^{i\delta} \right)M_+\nonumber\\
&& - \frac{{1 - {e^{ - 2i{\delta}}}}}{2}{M_{ee}},
\end{eqnarray}
we can derive that
\begin{eqnarray}
{M_{\mu \mu }} &=& \left( \frac{1}{c_{13}\tan 2\theta_{12}} - \frac{{{t_{13}}{e^{ - i{\delta}}}}}{{\sin 2{\theta_{23}}}} \right)M_- \nonumber \\
&& + \left( {\frac{{{e^{ - i{\delta}}}}}{{\tan 2{\theta_{13}}}} - \frac{1}{2}{t_{13}}{e^{i{\delta}}}} \right)M_+\nonumber\\
&& + \frac{{1 + {e^{ - 2i{\delta}}}}}{2}{M_{ee}} - M_{\mu\tau }^{\left( 0 \right)}\cos 2{\theta_{23}},
\nonumber\\
{M_{\tau \tau }} &=& \left( \frac{1}{c_{13}\tan 2\theta_{12}} + \frac{{{t_{13}}{e^{ - i{\delta}}}}}{{\sin 2{\theta_{23}}}} \right)M_- \nonumber \\
&& + \left( {\frac{{{e^{ - i{\delta}}}}}{{\tan 2{\theta_{13}}}} - \frac{1}{2}{t_{13}}{e^{i{\delta}}}} \right)M_+\nonumber\\
&& + \frac{{1 + {e^{ - 2i{\delta}}}}}{2}{M_{ee}} + M_{\mu\tau }^{\left( 0 \right)}\cos 2{\theta_{23}},
\nonumber\\
{M_{\mu \tau }} &=&  M_{\mu\tau }^{\left( 0 \right)}\sin 2{\theta_{23}},
\label{Eq:MainRelations}
\end{eqnarray}
and 
\begin{eqnarray}
{m_1}{e^{ - i{\phi _1}}} &=&  - \frac{t_{12}}{c_{13}}{M_ - } - t_{13}{e^{i{\delta}}}{M_ + } + {M_{ee}},
\nonumber\\
{m_2}{e^{ - i{\phi _2}}} &=& \frac{1}{c_{13}t_{12}}{M_ - } - t_{13}{e^{i{\delta}}}{M_ + } + {M_{ee}},
\nonumber\\
{m_3}{e^{ - i{\phi _3}}} &=& \frac{1}{t_{13}}{e^{ - i{\delta}}}{M_ + } + {e^{ - 2i{\delta}}}{M_{ee}}.
\label{Eq:MainRelationsMass}
\end{eqnarray}
Eqs.(\ref{Eq:MainRelations}) and (\ref{Eq:MainRelationsMass}) are the useful formulae which are announced by us in the introduction.

\subsection{Different parameterizations of $U$}
There are alternative parameterizations of the unitary matrix $U$. There are nine parametrizations of $U_0$ in the general case \cite{Fritzsch1998PRD}. Because the reactor mixing angle is small ($\sin^2\theta_{13}\ll 1$), the following four patterns (as well as the standard pattern $U_0^{\rm PDG}$ in Eq.(\ref{Eq:U_0})) are of great interest:
\begin{eqnarray}
U_0^{\rm I} &=& R_{23}(\theta_{23})R_{12}(\theta_{12})R_{13}(\theta_{13})
\nonumber \\
&\equiv&
\left( 
\begin{array}{*{20}{c}}
1 & 0 & 0 \\
0 & c_{23} & s_{23} \\
0 & -s_{23} & c_{23} \\
\end{array}
\right)
\left( 
\begin{array}{*{20}{c}}
c_{12} & s_{12} & 0 \\
-s_{12} & c_{12} & 0 \\
0 & 0 & 1 \\
\end{array}
\right)
\nonumber \\
&& \times
\left( 
\begin{array}{*{20}{c}}
c_{13} & 0 & s_{13}e^{-i\delta} \\
0 & 1 & 0 \\
-s_{13}e^{i\delta} & 0 & c_{13} \\
\end{array}
\right),
\label{Eq:U_0_I}
\end{eqnarray}
\begin{eqnarray}
U_0^{\rm II} &=& R_{23}(\theta_{23})R_{12}(\theta_{12})R_{23}(\theta_{13})
\nonumber \\
&\equiv&
\left( 
\begin{array}{*{20}{c}}
1 & 0 & 0 \\
0 & c_{23} & s_{23} \\
0 & -s_{23} & c_{23} \\
\end{array}
\right)
\left( 
\begin{array}{*{20}{c}}
c_{12} & s_{12} & 0 \\
-s_{12} & c_{12} & 0 \\
0 & 0 & 1 \\
\end{array}
\right)
\nonumber \\
&& \times
\left( 
\begin{array}{*{20}{c}}
1 & 0 & 0 \\
0 & c_{13} & s_{13}e^{-i\delta}\\
0 & -s_{13}e^{i\delta} & c_{13}\\
\end{array}
\right),
\label{Eq:U_0_II}
\end{eqnarray}
\begin{eqnarray}
U_0^{\rm III} &=& R_{13}(\theta_{13})R_{23}(\theta_{23})R_{12}(\theta_{12})
\nonumber \\
&\equiv&
\left( 
\begin{array}{*{20}{c}}
c_{13} & 0 & s_{13}e^{-i\delta}\\
0 & 1 & 0 \\
-s_{13}e^{i\delta} & 0 & c_{13}\\
\end{array}
\right)
\left( 
\begin{array}{*{20}{c}}
1 & 0 & 0 \\
0 & c_{23} & s_{23} \\
0 & -s_{23} & c_{23} \\
\end{array}
\right)
\nonumber \\
&& \times
\left( 
\begin{array}{*{20}{c}}
c_{12} & s_{12} & 0 \\
-s_{12} & c_{12} & 0 \\
0 & 0 & 1 \\
\end{array}
\right),
\label{Eq:U_0_III}
\end{eqnarray}
and
\begin{eqnarray}
U_0^{\rm IV} &=& R_{12}(\theta_{13})R_{23}(\theta_{23})R_{12}(\theta_{12})
\nonumber \\
&\equiv&
\left( 
\begin{array}{*{20}{c}}
c_{13} & s_{13}e^{-i\delta} & 0\\
-s_{13}e^{i\delta} & c_{13} & 0\\
0 & 0 & 1 \\
\end{array}
\right)
\left( 
\begin{array}{*{20}{c}}
1 & 0 & 0 \\
0 & c_{23} & s_{23} \\
0 & -s_{23} & c_{23} \\
\end{array}
\right)
\nonumber \\
&& \times
\left( 
\begin{array}{*{20}{c}}
c_{12} & s_{12} & 0 \\
-s_{12} & c_{12} & 0 \\
0 & 0 & 1 \\
\end{array}
\right).
\label{Eq:U_0_VI}
\end{eqnarray}

Since experimental analyses of neutrino oscillations are based on $U_0^{\rm PDG}K$, numerical values of the mixing angles other than $U_0^{\rm PDG}K$ need to be modified such as
\begin{eqnarray}
\sin^2 \theta^{\rm PDG}_{12}  &=&  \frac{\vert (U_0^{\rm I})_{12}\vert^2}{\vert (U_0^{\rm I})_{11}\vert^2+\vert (U_0^{\rm I})_{12}\vert^2},
\nonumber\\
\sin^2 \theta^{\rm PDG}_{23}  &=&  \frac{\vert (U_0^{\rm I})_{23}\vert^2}{\vert (U_0^{\rm I})_{23}\vert^2+\vert (U_0^{\rm I})_{33}\vert^2},
\nonumber\\
\sin^2 \theta^{\rm PDG}_{13}  &=&  \vert (U_0^{\rm I})_{13}\vert^2,
\label{Eq:modified_angles}
\end{eqnarray}
where $\theta^{\rm PDG}_{12,23,13}$ are $\theta_{12,23,13}$ in $U_0^{\rm PDG}$, which are directly compared with experimental data. 

It is possible to express $M_{\mu\mu, \mu\tau,\tau\tau}$ in terms of $M_{ee, e\mu,e\tau}$ in these different $U$'s.  As an example, in the case of $U_0^{\rm I}$, we find that
\begin{eqnarray}
{M_{\mu \mu }} &=& \left( {\frac{1}{{\tan 2{\theta _{12}}}} - \frac{1}{2}{t_{12}}{e^{ - 2i\delta }}} \right){M_ - }
\nonumber\\
&&
 + \left( {\frac{{{e^{ - i\delta }}}}{{{c_{12}}\tan 2{\theta _{13}}}} - \frac{{{t_{12}}}}{{\sin 2{\theta _{23}}}}} \right){M_ + }
\nonumber\\
&&
+ \frac{{1 + {e^{ - 2i\delta }}}}{2}{M_{ee}} - M_{\mu \tau }^{\left( 0 \right)}\cos 2{\theta _{23}},
\nonumber\\
{M_{\tau \tau }} &=& \left( {\frac{1}{{\tan 2{\theta _{12}}}} - \frac{1}{2}{t_{12}}{e^{ - 2i\delta }}} \right){M_ - }
\nonumber\\
&&
 + \left( {\frac{{{e^{ - i\delta }}}}{{{c_{12}}\tan 2{\theta _{13}}}} + \frac{{{t_{12}}}}{{\sin 2{\theta _{23}}}}} \right){M_ + }
\nonumber\\
&&
 + \frac{{1 + {e^{ - 2i\delta }}}}{2}{M_{ee}} + M_{\mu \tau }^{\left( 0 \right)}\cos 2{\theta _{23}},
\nonumber\\
{M_{\mu \tau }} &=& M_{\mu \tau }^{\left( 0 \right)}\sin 2{\theta _{23}},
\label{Eq:MainRelations-2}
\end{eqnarray}
where
\begin{eqnarray}
M_{\mu \tau }^{\left( 0 \right)} &=&  - \left( {\frac{1}{{\tan 2{\theta _{12}}}} + \frac{1}{2}{t_{12}}{e^{ - 2i\delta }}} \right){M_ - }
\nonumber\\
&&
 + \left( {\frac{{{e^{ - i\delta }}}}{{{c_{12}}\tan 2{\theta _{13}}}} - \frac{{{t_{12}}}}{{\tan 2{\theta _{23}}}}} \right){M_ + }
\nonumber\\
&&
 - \frac{{1 - {e^{ - 2i\delta }}}}{2}{M_{ee}}.
\label{Eq:Mmutau(0)variation}
\end{eqnarray}
Roughly speaking, except for the sign of $\delta$, Eq.(\ref{Eq:MainRelations-2}) is obtained from Eq.(\ref{Eq:MainRelations}) by the replacement of $M_{ij}\rightarrow e^{i\delta}M_{ij}$ ($i,j=\mu,\tau$), $M_+\leftrightarrow M_-$ and $\theta_{12}\leftrightarrow \theta_{13}$.  The replacement of $\theta_{12}\leftrightarrow \theta_{13}$ can be understood as the effect of the interchanged rotations between $R_{13}(\theta_{13})$ and $R_{12}(\theta_{12})$ to get $U_0^{\rm I}$ from $U_0^{\rm PDG}$. We have also numerically evaluated $\sin^2\theta_{12,23,13}$ and $\delta$ and have observed that $\sin^2\theta_{12,23}$ and $\delta$ indicate no significant deviation due to the smallness of $\sin^2\theta_{13}$. On the other hand, $\vert (U_0^{\rm I})_{13}\vert = c_{12}s_{13}$ requires larger values of $s_{13}$ to yield the same magnitude of $\sin^2\theta^{\rm PDG}_{13}$ owing to the suppression by $c^2_{12}$.  The neutrino masses turn out to be:
\begin{eqnarray}
{m_1}{e^{ - i{\phi _1}}} &=&  - {t_{12}}{M_ - }
\nonumber\\
&& 
+ \left[ \begin{array}{l}
\frac{1}{2}\left( {\frac{1}{2} - {e^{2i{\delta}}}} \right){t_{12}}{t_{23}}
-\frac{{{t_{13}}}}{{{c_{12}}}}{e^{i{\delta}}}
\end{array} \right]{M_ + }
\nonumber\\
&&
 + {M_{ee}},
\nonumber\\
{m_2}{e^{ - i{\phi _2}}} &=&  - {t_{12}}{M_ - } + \frac{{{t_{23}}}}{{2{c_{12}}}}{M_ + } + {M_{ee}},
\nonumber\\
{m_3}{e^{ - i{\phi _3}}} &=&  - {t_{12}}{e^{ - 2i{\delta}}}{M_ - }
\nonumber\\
&&
 + \left[ \begin{array}{l}
\frac{1}{2}\left( {\frac{1}{2}{e^{ - 2i{\delta}}} - 1} \right){t_{12}}{t_{23}}
 + 
\frac{1}{{{t_{13}}{c_{12}}}}{e^{ - i{\delta}}}
\end{array} \right]{M_ + }
\nonumber\\
&&
 + {e^{ - 2i{\delta}}{M_{ee}}}.
\label{Eq:MainRelationsMassvariation}
\end{eqnarray}
There is no simple relationship between Eq.(\ref{Eq:MainRelationsMass}) and Eq.(\ref{Eq:MainRelationsMassvariation}) realized by replacement of the masses and the mixing angles. 

It will be interesting to discuss the phase structure of $M$ determined by these different $U$'s because there appear to be the different dependence of $\delta$ from the one derived by $U^{\rm PDG}_0$. Since predictions of $m_{1,2,3}$ in various $U$'s are expected to be quite different from those by $U^{\rm PDG}_0$ as have been learned from the comparison between  Eq.(\ref{Eq:MainRelationsMass}) and Eq.(\ref{Eq:MainRelationsMassvariation}), the Majorana phases may give different behaviors from those predicted by $U^{\rm PDG}_0$. However, to discuss experimental allowed regions for CP-violating phases including $\delta$ as well as $\theta_{12,23,13}$ in $U_0^{\rm I, II, III, IV}$ is beyond the scope of the present discussions.

Moreover, there is another parameterization induced by three Dirac CP-violating phases $\gamma_{23}, \gamma_{13},\gamma_{12}$ without explicitly referring to Majorana phases \cite{Rodejohann2011PRD}: 
\begin{eqnarray}
U &=&
\left( 
\begin{array}{*{20}{c}}
1 & 0 & 0 \\
0 & c_{23} & \tilde{s}_{23} \\
0 & -\tilde{s}_{23}^\ast & c_{23} \\
\end{array}
\right)
\left( 
\begin{array}{*{20}{c}}
c_{13} & 0 & \tilde{s}_{13} \\
0 & 1 & 0 \\
-\tilde{s}_{13}^\ast & 0 & c_{13} \\
\end{array}
\right)
\nonumber \\
&& \times
\left( 
\begin{array}{*{20}{c}}
c_{12} & \tilde{s}_{12} & 0 \\
-\tilde{s}_{12}^\ast & c_{12} & 0 \\
0 & 0 & 1 \\
\end{array}
\right),
\label{Eq:tildeU}
\end{eqnarray}
where $\tilde{s}_{ij}=s_{ij}e^{-i\gamma_{ij}}$. All three Dirac CP-violating phases $\gamma_{23,13,12}$ are physical. These Dirac CP-violating phases are related to the phases in the PDG version as $\gamma_{13} = \delta - \alpha_3/2,$ $\gamma_{12}=\alpha_2/2$ and  $\gamma_{23}=(\alpha_3-\alpha_2)/2$ \cite{Yasue2012PLB}. The advantages of choosing $\gamma_{23, 13, 12}$ instead of $\delta, \alpha_2, \alpha_3$ have been discussed in Refs.\cite{Rodejohann2011PRD,Yasue2012PLB}. For an example, this alternative parameterization provides that only the two Majorana phases appear in the effective Majorana neutrino mass for the neutrino less double beta decay.  Since constraints on $\delta$ and $\alpha_2, \alpha_3$ to be derived in Sec.\ref{sec:texture_two_zeros} can be readily converted to those on $\gamma$'s, we use the PDG parameterisation to compare our results with conclusions obtained from other papers.

\section{\label{sec:texture_two_zeros}Texture two zeros}

\subsection{Texture two zeros}
Texture two zeros for the flavour neutrino mass matrix $M_{ij}$ ($i,j=e,\mu,\tau$) have been extensively studied in literature (see for examples \cite{Fritzsch2011JHEP, Meloni2014PRD,Dev2014PRD,Zhou2015arXiv}). 
There are 15 possible combination of two vanishing independent elements in the flavour neutrino mass matrix. If we require a non-vanishing Majorana effective mass for the neutrino less double beta decay and maximal or approximately maximal CP violation, the interesting textures are the following only four \cite{Meloni2014PRD,Dev2014PRD} (or less \cite{Zhou2015arXiv})
\begin{eqnarray}
&& B_1:
\left( 
\begin{array}{*{20}{c}}
M_{ee} & M_{e\mu} & 0 \\
M_{e\mu} & 0 & M_{\mu\tau} \\
0 & M_{\mu\tau} & M_{\tau\tau} \\
\end{array}
\right),
\nonumber \\
&& B_2:
\left( 
\begin{array}{*{20}{c}}
M_{ee} & 0 & M_{e\tau} \\
0& M_{\mu\mu} & M_{\mu\tau} \\
M_{e\tau} & M_{\mu\tau} & 0 \\
\end{array}
\right),
\nonumber \\
&&B_3:
\left( 
\begin{array}{*{20}{c}}
M_{ee} & 0 & M_{e\tau} \\
0 & 0 & M_{\mu\tau} \\
M_{e\tau} & M_{\mu\tau} & M_{\tau\tau} \\
\end{array}
\right),
\nonumber \\
&& B_4:
\left( 
\begin{array}{*{20}{c}}
M_{ee} & M_{e\mu} & 0 \\
M_{e\mu} & M_{\mu\mu} & M_{\mu\tau} \\
0 & M_{\mu\tau} & 0 \\
\end{array}
\right).
\label{Eq:B1B2B3B4}
\end{eqnarray}

The texture zeros in the Majorana neutrino mass matrix can be realized by imposing appropriate symmetries on the Lagrangian of neutrino mass models such as an Abelian flavour symmetry (e.g., the cyclic group $Z_n$) \cite{Dev2014PRD,textureZerosAndAbelianSymmetries} as well as a non-Abelian symmetry \cite{Zhou2015arXiv}. The stability of texture zeros with respect to higher order corrections is protected by these symmetries \cite{Dev2014PRD,textureZerosAndStability}.

For example, $B_1, B_2, B_3$ and $B_4$ textures are obtained by imposing the discrete cyclic group $Z_3$ \cite{Dev2014PRD} in the framework of type-I \cite{TypeISeesaw} plus type-II \cite{TypeIISeesaw} seesaw mechanism. There are other symmetry realization of $B_1, B_2, B_3$ and $B_4$ textures. The $B_1$ and $B_2$ textures have been realized by using $A_4$ or its $Z_3$ subgroup \cite{Hirsch2007PRL}. The $B_3$ and $B_4$ textures have been obtained by softly breaking the $L_\mu-L_\tau$ symmetry \cite{Rodejohann2005PAN}.

\subsection{Analytical solutions}

The mass matrices based on the texture two zeros are good examples to demonstrate the usefulness of the formulae in Eq.(\ref{Eq:MainRelations}) and Eq.(\ref{Eq:MainRelationsMass}). For $B_1$ texture, because of $M_{e\tau}=M_{\mu\mu}=0$, it is evident that 
\begin{eqnarray}
M_a &=& f_a(\theta_{12},\theta_{23}, \theta_{13},\delta)M_{ee},\\
m_je^{-i\phi_j} &=& f_j(\theta_{12},\theta_{23}, \theta_{13},\delta)M_{ee}, \label{Eq:m_f_Mee}
\end{eqnarray}
where $a=e\mu,\mu\tau, \tau\tau$ and $j=1,2,3$. The non-vanishing flavour neutrino masses $M_{e\mu},M_{\mu\tau}, M_{\tau\tau}$ and mass eigenstates $m_1,m_2,m_3$ are the functions of only $M_{ee}$ (with mixing angles and phases fixed). More concretely, with the following definitions in mind,
\begin{eqnarray}
A_1&=&c_{23}^2+s_{23}^2 e^{-2i\delta}, \nonumber \\
A_2&=&s_{23}^2+c_{23}^2 e^{-2i\delta}, \nonumber \\
B_1&=& \frac{2c_{23}^2}{c_{13}\tan 2\theta_{12}} - t_{13}\sin 2\theta_{23} e^{-i\delta}, \nonumber \\
B_2&=& \frac{2s_{23}^2}{c_{13}\tan 2\theta_{12}} + t_{13}\sin 2\theta_{23} e^{-i\delta}, \nonumber \\
B_3&=&  \frac{\sin 2\theta_{23}}{c_{13}\tan 2\theta_{12}} +t_{13} \cos 2\theta_{23} e^{-i\delta}, \nonumber \\
C_1&=& 2  \left( \frac{s_{23}^2 e^{-i\delta}}{\tan 2\theta_{13}} -\frac{t_{13} c_{23}^2 e^{i\delta}}{2} \right), \nonumber \\
C_2&=& 2  \left( \frac{c_{23}^2 e^{-i\delta}}{\tan 2\theta_{13}} -\frac{t_{13} s_{23}^2 e^{i\delta}}{2} \right), \nonumber \\
C_3&=&  \sin 2\theta_{23} \left( \frac{e^{-i\delta}}{\tan 2\theta_{13}} +\frac{t_{13} e^{i\delta}}{2} \right), 
\end{eqnarray}
we obtain  
\begin{eqnarray}
M_{e\mu}&=&-\frac{A_1}{c_{23}B_1 + s_{23}C_1}M_{ee},
\nonumber \\
M_{\mu\tau}&=&\left(-A_1\frac{-c_{23}B_3 + s_{23}C_3}{c_{23}B_1 + s_{23}C_1} - \frac{1-e^{-2i\delta}}{2} \sin 2\theta_{23}\right) M_{ee},
\nonumber \\
M_{\tau\tau}&=&\left(-A_1\frac{c_{23}B_2+ s_{23}C_2}{c_{23}B_1 + s_{23}C_1} + A_2 \right) M_{ee},
\label{Eq:M_B1}
\end{eqnarray}
and
\begin{eqnarray}
m_1 e^{-i \phi_1}&=&\left (A_1\frac{ \frac{t_{12}c_{23}}{c_{13}} +t_{13}s_{23}e^{i\delta} }{c_{23}B_1 + s_{23}C_1} + 1 \right) M_{ee},
\nonumber \\
m_2 e^{-i \phi_2}&=&\left (A_1\frac{ -\frac{c_{23}}{c_{13}t_{12}} +t_{13}s_{23}e^{i\delta} }{c_{23}B_1 + s_{23}C_1} + 1 \right) M_{ee},
\nonumber \\
m_3 e^{-i \phi_3}&=&\left (A_1\frac{ -\frac{s_{23}}{t_{13}}e^{-i\delta} }{c_{23}B_1 + s_{23}C_1} + e^{-2i\delta} \right) M_{ee},
\label{Eq:m_B1}
\end{eqnarray}
for $B_1$ texture. 

Similarly, the non-vanishing flavour neutrino masses as well as mass eigenstates for $B_2, B_3$ and $B_4$ textures are expressed in terms of $M_{ee}$ as 
\begin{eqnarray}
M_{e\tau}&=&\frac{A_2}{s_{23}B_2 - c_{23}C_2}M_{ee},
\nonumber \\
M_{\mu\mu}&=&\left(A_2\frac{-s_{23}B_1+ c_{23}C_1}{s_{23}B_2 - c_{23}C_2} + A_1 \right) M_{ee},
\\
M_{\mu\tau}&=&\left(A_2\frac{s_{23}B_3 + c_{23}C_3 }{s_{23}B_2 - c_{23}C_2} - \frac{1-e^{-2i\delta}}{2} \sin 2\theta_{23}\right) M_{ee},
\nonumber 
\label{Eq:M_B2}
\end{eqnarray}
and
\begin{eqnarray}
m_1 e^{-i \phi_1}&=&\left (A_2\frac{ \frac{t_{12}s_{23}}{c_{13}} -t_{13}c_{23}e^{i\delta} }{s_{23}B_2 - c_{23}C_2} + 1 \right) M_{ee},
\nonumber \\
m_2 e^{-i \phi_2}&=&\left (A_2\frac{ -\frac{s_{23}}{c_{13}t_{12}} -t_{13}c_{23}e^{i\delta} }{s_{23}B_2 - c_{23}C_2} + 1 \right) M_{ee},
\nonumber \\
m_3 e^{-i \phi_3}&=&\left (A_2\frac{ \frac{c_{23}}{t_{13}}e^{-i\delta} }{s_{23}B_2 - c_{23}C_2} + e^{-2i\delta} \right) M_{ee},
\label{Eq:m_B2}
\end{eqnarray}
for $B_2$ texture,
\begin{eqnarray}
M_{e\tau}&=&\frac{A_1}{s_{23}B_1 - c_{23}C_1}M_{ee},
\nonumber \\
M_{\mu\tau}&=&\left(A_1\frac{s_{23}B_3 + c_{23}C_3 }{s_{23}B_1 - c_{23}C_1} - \frac{1-e^{-2i\delta}}{2} \sin 2\theta_{23}\right) M_{ee},
\nonumber \\
M_{\tau\tau}&=&2\left(A_1\frac{(s_{23}B_3+ c_{23}C_3)\cos 2\theta_{23} - s_{23}t_{13}e^{-i\delta}}{\sin 2\theta_{23}(s_{23}B_1 - c_{23}C_1)} \right.
\nonumber \\
&& \left. - \frac{1-e^{-2i\delta}}{2} \cos 2\theta_{23} \right) M_{ee},
\label{Eq:M_B3}
\end{eqnarray}
and
\begin{eqnarray}
m_1 e^{-i \phi_1}&=&\left (A_1\frac{ \frac{t_{12}s_{23}}{c_{13}} -t_{13}c_{23}e^{i\delta} }{s_{23}B_1 - c_{23}C_1} + 1 \right) M_{ee},
\nonumber \\
m_2 e^{-i \phi_2}&=&\left (A_1\frac{ -\frac{s_{23}}{c_{13}t_{12}} -t_{13}c_{23}e^{i\delta} }{s_{23}B_1 - c_{23}C_1} + 1 \right) M_{ee},
\nonumber \\
m_3 e^{-i \phi_3}&=&\left (A_1\frac{ \frac{c_{23}}{t_{13}}e^{-i\delta} }{s_{23}B_1 - c_{23}C_1} + e^{-2i\delta} \right) M_{ee},
\label{Eq:m_B3}
\end{eqnarray}
for $B_3$ texture and 
\begin{eqnarray}
M_{e\mu}&=&-\frac{A_2}{c_{23}B_2 + s_{23}C_2}M_{ee},
\nonumber \\
M_{\mu\mu}&=&\left(-A_2\frac{c_{23}B_1+ s_{23}C_1}{c_{23}B_2 + s_{23}C_2} + A_1 \right) M_{ee},
\nonumber \\
M_{\mu\tau}&=&\left(-A_2\frac{-c_{23}B_3 + s_{23}C_3 }{c_{23}B_2 + s_{23}C_2} - \frac{1-e^{-2i\delta}}{2} \sin 2\theta_{23}\right) M_{ee},
\nonumber \\
\label{Eq:M_B4}
\end{eqnarray}
and
\begin{eqnarray}
m_1 e^{-i \phi_1}&=&\left (A_2\frac{ \frac{t_{12}c_{23}}{c_{13}} + t_{13}s_{23}e^{i\delta} }{c_{23}B_2 + s_{23}C_2} + 1 \right) M_{ee},
\nonumber \\
m_2 e^{-i \phi_2}&=&\left (-A_2\frac{ \frac{c_{23}}{c_{13}t_{12}} -t_{13}s_{23}e^{i\delta} }{c_{23}B_2 + s_{23}C_2} + 1 \right) M_{ee},
\nonumber \\
m_3 e^{-i \phi_3}&=&\left (-A_2\frac{ \frac{s_{23}}{t_{13}}e^{-i\delta} }{c_{23}B_2 + s_{23}C_2} + e^{-2i\delta} \right) M_{ee},
\label{Eq:m_B4}
\nonumber \\
\end{eqnarray}
for $B_4$ texture.

The physical CP-violating Majorana phases $\alpha_2,\alpha_3$ depend on the CP-violating Dirac phase $\delta$ in all cases of $B_1,B_2,B_3$ and $B_4$. Since the mass eigenstate $m_j$ is obtained as Eq.(\ref{Eq:m_f_Mee}), the Majorana phase $\phi_j$ depends on not only $\delta$ but also ${\rm arg} (M_{ee})$ as follows;
\begin{eqnarray}
\phi_j = -{\rm arg}(m_j e^{-i \phi_j}) = -{\rm arg} (f_j(\delta)) - {\rm arg} (M_{ee}).
\end{eqnarray}
On the contrary, the physical CP-violating Majorana phase $\alpha_j$ is specified by two combinations made of $\phi_1,\phi_2,\phi_3$ such as  
\begin{eqnarray}
\alpha_j = \phi_j - \phi_1 = {\rm arg}\left( \frac{f_1(\delta)}{f_j(\delta)} \right),
\label{Eq:MajoranaPhases}
\end{eqnarray}
which depends on the CP-violating Dirac phase $\delta$. For example, the function $f_1$, $f_2$ and $f_3$ are obtained from Eq.(\ref{Eq:m_B1}) as
\begin{eqnarray}
f_1 &=& A_1\frac{ \frac{t_{12}c_{23}}{c_{13}} +t_{13}s_{23}e^{i\delta} }{c_{23}B_1 + s_{23}C_1} + 1,
\nonumber \\
f_2 &=&  A_1\frac{ -\frac{c_{23}}{c_{13}t_{12}} +t_{13}s_{23}e^{i\delta} }{c_{23}B_1 + s_{23}C_1} + 1,
\nonumber \\
f_3 &=& A_1\frac{ -\frac{s_{23}}{t_{13}}e^{-i\delta} }{c_{23}B_1 + s_{23}C_1} + e^{-2i\delta},
\label{Eq:f1_f2_f3_B1}
\end{eqnarray}
for $B_1$ texture. The function of $f_j$ for other textures is immediately obtained from Eqs.(\ref{Eq:m_B2}), (\ref{Eq:m_B3}) and (\ref{Eq:m_B4}). 

The analytical expressions for the flavour neutrino masses, mass eigenstates and the physical CP-violating Majorana phases are obtained exactly. We observed the flavour neutrino masses and mass eigenstates for $B_1,B_2,B_3$ and $B_4$ depend on $M_{ee}$ (with mixing angles and phases fixed). Also, the physical CP-violating Majorana phases depend on the CP-violating Dirac phase $\delta$. These clear understanding is obtained from the observation based on the formulae in Eqs.(\ref{Eq:MainRelations}) and (\ref{Eq:MainRelationsMass}).

\subsection{Numerical calculations}
\begin{table}[t]
\caption{Octant of $\theta_{23}$ for $B_1,B_2,B_3$ and $B_4$ textures ($\ell$: lower octant $\theta_{23}<45^\circ$, $u$: upper octant $\theta_{23}>45^\circ$) \cite{Meloni2014PRD,Dev2014PRD}. }
\begin{center}
\begin{tabular}{c|cccc}
\hline
Ordering     & $B_1$ & $B_2$ & $B_3$ & $B_4$ \\
\hline
NO & $\ell$ & $u$ & $\ell$ & $u$ \\
IO & $u$ & $\ell$ & $u$ & $\ell$ \\
\hline
\end{tabular}
\end{center}
\label{TABLE:octant}
\end{table}

Although, we have reached our main goal of this paper to see the usefulness of the formulae, some numerical and semi-analytical calculations may be required to confirm results of our discussions. 

First, we show the consistency between our results and conclusions obtained from other papers. The results from the precise parameter search for $M_{ij}$ [eV] are reported by Dev et al. \cite{Dev2014PRD}. For $B_1$ texture, they obtained
\begin{eqnarray}
M_{e\mu}&=&0.00018+0.00258i,
\nonumber \\
M_{\mu\tau}&=&-0.11001-0.00287i,
\nonumber \\
M_{\tau\tau}&=&-0.01105-0.00024i,
\end{eqnarray}
for $M_{ee}=0.10479+0.00127i$ with $s^2_{12} = 0.32$, $s^2_{23} = 0.476$, $s^2_{13} = 0.0246$ and $\delta= -90.65^\circ$. The predicted values from our exact expression in Eq.(\ref{Eq:M_B1}) are
\begin{eqnarray}
M_{e\mu}&=&0.000185+0.00258i,
\nonumber \\
M_{\mu\tau}&=&-0.110-0.00287i,
\nonumber \\
M_{\tau\tau}&=&-0.0110-0.000237i,
\end{eqnarray}
for the same values of $M_{ee}, s^2_{12},s^2_{23},s^2_{13}$ and $\delta$. The mass eigenstates $(m_1,m_2,m_3)$ are estimated by Meloni et al. \cite{Meloni2014PRD} and, for $B_1$ texture, they reported
\begin{eqnarray}
(m_1,m_2,m_3)=(6.27, 6.33,7.97) \times 10^{-2} {\rm eV},
\end{eqnarray}
for $M_{ee}=6.32\times 10^{-2}$ eV with $s^2_{12} = 0.306$, $s^2_{23} = 0.446$, $s^2_{13} = 0.0229$, $\delta= -91.7^\circ$, $\alpha_2=5.61^\circ$ and $\alpha_3=183.8^\circ$. We have obtained the same masses from our exact expression in Eq.(\ref{Eq:m_B1}). The similar consistency for the remaining textures, $B_2, B_3$ and $B_4$ is also guaranteed.

%
\begin{figure}[t]
\begin{center}
\includegraphics[width=8.0cm]{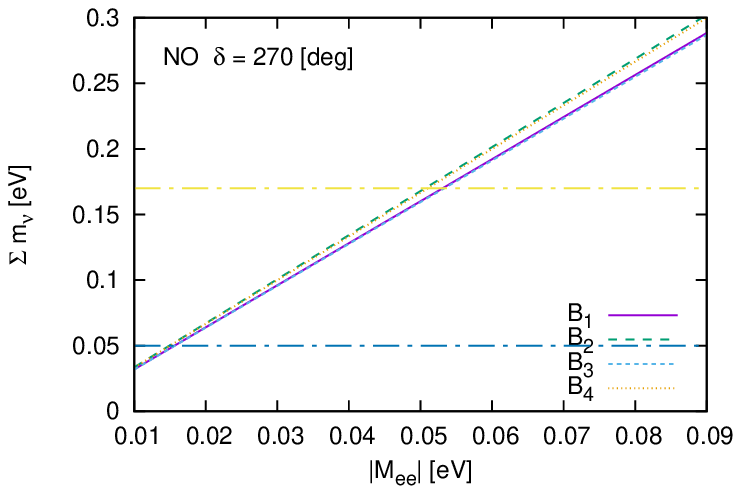}
\includegraphics[width=8.0cm]{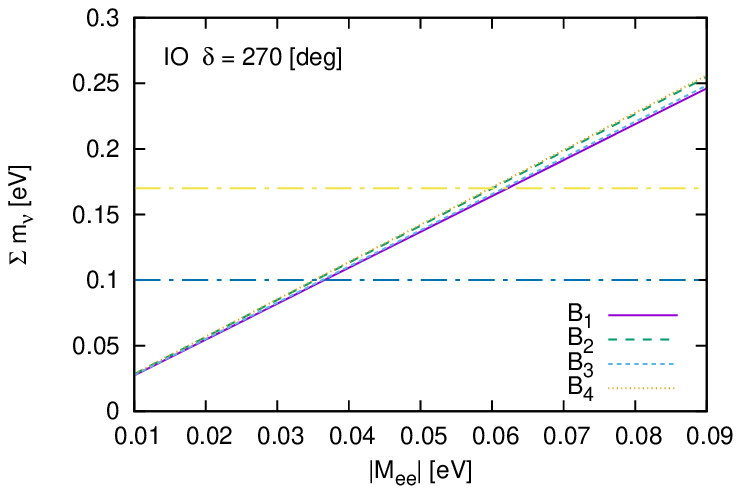}
\caption{$\sum m_\nu$ - $\vert M_{ee}\vert$ planes for $\delta = 270^\circ$. The upper panel shows the relation for the normal ordering (NO) while the lower panel shows the relation for inverted ordering (IO). The upper (lower) horizontal dashed-and-dotted lines in the figures show the observed upper (lower) limit of the sum of the light neutrino masses from the recent results of the Planck  (neutrino oscillations) experiments.}
\label{fig:m_mee}
\end{center}
\end{figure}
%
\begin{figure*}[t]
\begin{center}
\includegraphics[width=8.0cm]{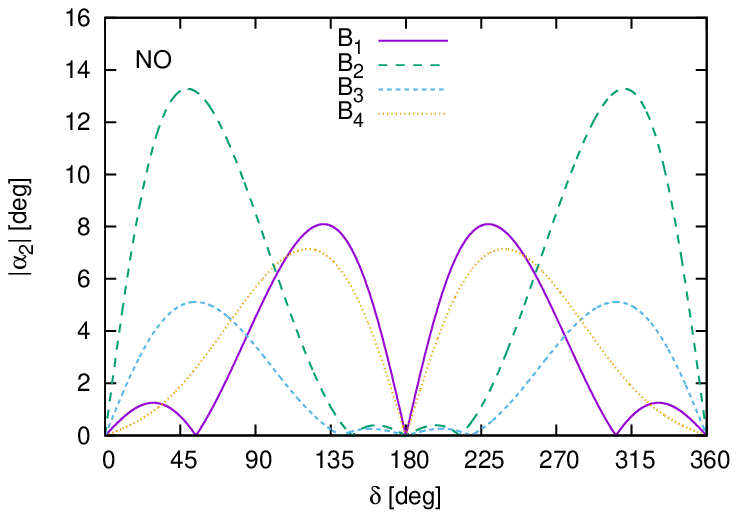}
\includegraphics[width=8.0cm]{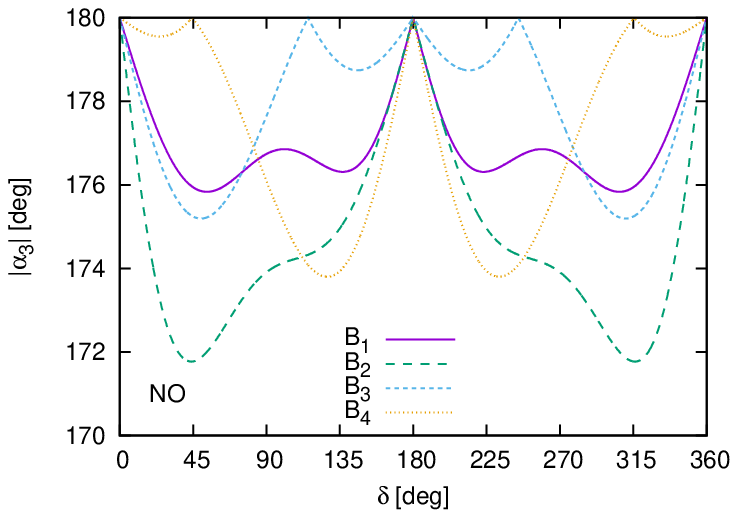}
\includegraphics[width=8.0cm]{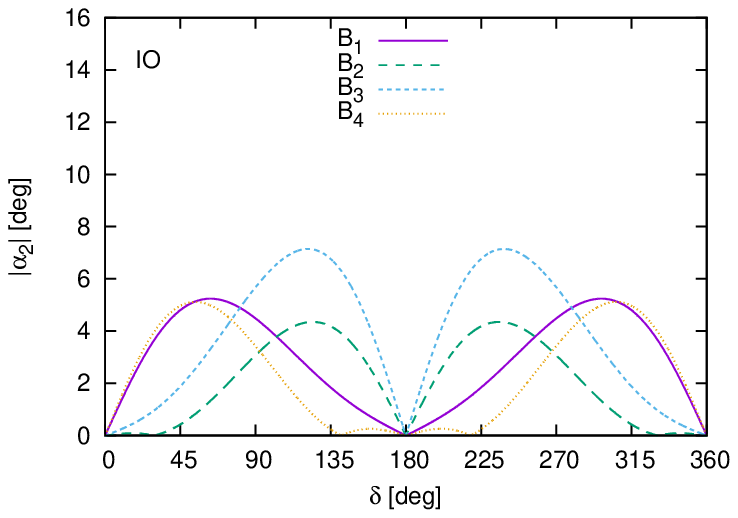}
\includegraphics[width=8.0cm]{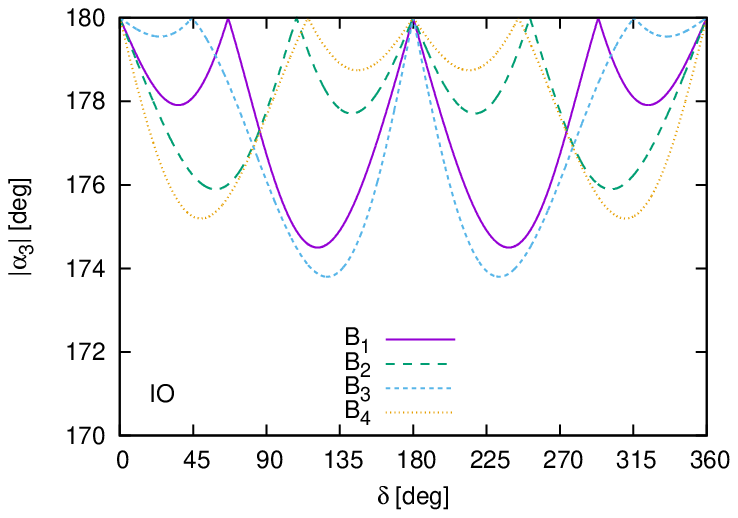}
\caption{The dependence of the Majorana CP-violating phases $\alpha_2$ and $\alpha_3$ on the CP-violating Dirac phase $\delta$. The upper two panels show the magnitude of $\alpha_2$ (left) and $\alpha_3$ (right) for the normal ordering (NO) while the lower two panels show the magnitude of $\alpha_2$ (left) and $\alpha_3$ (right) for the inverted ordering (IO).  }
\label{fig:phase}
\end{center}
\end{figure*}

Second, we estimate the upper limit of $\vert M_{ee} \vert$ by the expression in Eq.(\ref{Eq:m_f_Mee}) with observed data of $\sum m_{\nu}=\sum_j \vert m_j e^{-i \phi_j} \vert$. To calculate $f_j(\theta_{12},\theta_{23},\theta_{13},\delta)$, we use the following values of three mixing angles \cite{Nudata}
\begin{eqnarray}
\sin^2 \theta_{12} &=& 0.304,
\nonumber\\
\sin^2 \theta_{23} &=& 0.452 \ {\rm or} \ 0.579,
\nonumber\\
\sin^2 \theta_{13} &=& 0.0219.
\label{Eq:observedangle}
\end{eqnarray}
The octant of $\theta_{23}$ (i.e. lower octant $\theta_{23}<45^\circ$ or upper octant $\theta_{23}>45^\circ$) is still unresolved problem.   According to Dev et al. \cite{Dev2014PRD} and Meloni et al. \cite{Meloni2014PRD}, we employ the values of $\theta_{23}$ given for both octant which are shown in TABLE \ref{TABLE:octant}. For the CP-violating Dirac phase, the T2K collaboration has reported the result of their new analysis \cite{T2K2015PRD} and has shown that the likelihood maximum is reached at $\delta=270^\circ$ and $\sin^2 \theta_{23}=0.528$. 

FIG. \ref{fig:m_mee} shows the $\sum m_\nu$ - $\vert M_{ee}\vert$ planes for $\delta = 270^\circ$. The upper panel shows the relation for the normal ordering (NO), $m_1 < m_2 < m_3$, while the lower panel shows the relation for the inverted ordering (IO), $m_3 < m_1 < m_2$. 
The upper (lower) horizontal dashed-and-dotted lines in the figures show the observed upper (lower) limit of the sum of the light neutrino masses $\sum m_\nu$ from the recent results of the Planck \cite{Planck2015arXiv} (neutrino oscillations \cite{Nudata}) experiments. The upper limit is $\sum m_\nu \le 0.17$ eV. The lower limits are $\sum m_\nu \gtrsim m_1 + \Delta m_{21} + \Delta m_{31} \cong 0.05$ eV for NO (with $m_1=0$) and $\sum m_\nu \gtrsim 2\vert \Delta m_{32} \vert+ m_3 \cong 0.1$ eV for IO (with $m_3=0$) where we have used the values of the squared mass differences,  $\Delta m_{ij}^2 \equiv m_i^2-m_j^2$, as \cite{Nudata}  
\begin{eqnarray}
\Delta m_{21}^2 &=&7.5\times 10^{-5} {\rm eV}^2 \ {\rm for \ NO \ and \ IO},
\nonumber \\
\Delta m_{31}^2 &=& 2.46\times 10^{-3} {\rm eV}^2 \ {\rm for \ NO},
\nonumber \\
\Delta m_{32}^2 &=& -2.45\times 10^{-3} {\rm eV}^2 \ {\rm for \ IO}.
\end{eqnarray}

The lines for $B_1$ and $B_3$ as well as $B_2$ and $B_4$ are almost overlapped each other because the neutrino mass spectrum is to be quasi-degenerate as $m_1\simeq m_2 \simeq m_3 t^2_{23}$ for $B_1,B_3$ or $m_1\simeq m_2 \simeq m_3 /t^2_{23}$ for $B_2,B_4$ \cite{Fritzsch2011JHEP,Meloni2014PRD}. The upper limit of $\vert M_{ee} \vert$ is around $0.05$ eV (NO) or $0.06$ eV (IO) for maximal CP-violation $\delta = 270^\circ$. These upper limits are consistent with the result in Ref.\cite{Meloni2014PRD}. 

The theoretically expected half-life of the neutrino less double beta decay is proportional to the effective Majorana neutrino mass $m_{\beta\beta}$ where $\vert m_{\beta\beta}\vert =\vert M_{ee}\vert = \vert \sum_{i=1}^3 U_{ei}^2 m_i \vert$. The estimated magnitude of the effective Majorana mass from the experiments is $\vert M_{ee} \vert = 0.20-2.5$ eV \cite{Cremonesi2013arXiv}. In the future experiments, a desired sensitivity $\vert M_{ee} \vert \simeq$ a few $10^{-2}$ eV will be reached \cite{Benato2015arXiv}.

Finally, we study the dependence of the CP-violating Majorana phases $\alpha_2$ and $\alpha_3$ on the CP-violating Dirac phase $\delta$. In FIG.\ref{fig:phase}, the upper two panels show the magnitude of $\alpha_2$ (left) and $\alpha_3$ (right) for the normal ordering (NO) while the lower two panels show the magnitude of $\alpha_2$ (left) and $\alpha_3$ (right) for the inverted ordering (IO). The behaviour of the curves in FIG.\ref{fig:phase} can be understand by the semi-analytical expressions. For example, we obtain
\begin{eqnarray}
\frac{f_1}{f_2} &\simeq& -2.29+\frac{1}{0.287-0.0421 e^{i\delta}}+\frac{1}{-0.611+2.20 e^{i\delta}},
\nonumber \\
\frac{f_1}{f_3} &\simeq& - 0.797-0.247 e^{i\delta} \nonumber \\
 &&+ \frac{1}{-0.669-5.43 e^{i\delta} + (0.512-1.91 e^{i\delta})^{-1} },
\end{eqnarray}
with the mixing angles in Eq.(\ref{Eq:observedangle}) for $B_1$ texture in the case of NO (we use lower octant of $\theta_{23}$) and $f_1/f_2=2.42$ and $f_1/f_3=-1.19$ for $\delta=0^\circ$ yield $\alpha_2=0^\circ$ and $\alpha_3=180^\circ$, respectively. 

\section{\label{sec:summary}Summary}
We have demonstrated the usefulness of flavour neutrino masses expressed in terms of $M_{ee},M_{e\mu}$ and $M_{e\tau}$. The analytical expressions for the flavour neutrino masses and mass eigenstates for texture two zeros are obtained exactly. These flavour neutrino masses and mass eigenstates depend on $M_{ee}$ (with mixing angles and phases fixed). The analytical expressions for the physical CP-violating Majorana phases are also obtained exactly. These phases depend on the CP-violating Dirac phase. 

To confirm results of our discussions more concretely, we have discussed the following three subjects based on numerical and semi-analytical calculations: (1) the consistency between our results and conclusions obtained from other papers, (2) the upper limit of the effective Majorana neutrino mass for the neutrino less double beta decay for maximal Dirac CP-violation and (3) the dependence of the CP-violating Majorana phases on the CP-violating Dirac phase. 



\end{document}